\documentstyle[12pt,aaspp4]{article}

\def\kms{\hbox {\ km\ s$^{-1}\,$}}
\def\hi {H{\small I} }
\def\hii {H{\small II} }
\def\Msun{\hbox{M$_{\odot}$}}
\def\g132{G132.6--0.7--25.3}

\lefthead{Normandeau}
\righthead{The \hi shell G132.6--0.7--25.3}

\begin{document}

\title{The HI shell G132.6--0.7--25.3: A Supernova Remnant or an Old
Wind-Blown Bubble?} 
\author{Magdalen Normandeau}
\affil{Astronomy Department, University of California,
    Berkeley, CA 94720-3411, United States}
\author{A.\ R.\ Taylor}
\affil{Department of Physics and Astronomy, University of Calgary,
    Calgary AB\ \ T2N 1N4, Canada}
\author{P.\ E.\ Dewdney}
\affil{Dominion Radio Astrophysical Observatory, Box 248,
    Penticton BC\ \ V2A 6K3, Canada}
\author{Shantanu Basu}
\affil{Department of Physics and Astronomy, University of Western
    Ontario, London ON\ \ N6A 3K7, Canada}

\begin{abstract}
Data from the Canadian Galactic Plane Survey
reveal an abundance of \hi shells and arcs in the disk of our
galaxy. While their shape is suggestive of stellar winds or supernovae
influence, very few of these structures have been examined in detail
thus far. A fine example is 
an \hi shell in the outer Galaxy with no continuum counterpart
discovered in the survey's pilot project.
Its size and kinematics suggest that it was created by the winds of a
single late-type O star which has since evolved off the main
sequence or by a supernova explosion. A B1 Ia star at the centre of
the shell, in projection, is a possible candidate for energy
source if the shell is assumed to be wind-blown. 
The shell's shape implies a surprisingly small scale height of less
than about 30 pc for the surrounding gas if the elongation is due to
evolution in a density gradient.  

\end{abstract}

\keywords{ISM: Bubbles -- ISM: HI -- ISM: Kinematics and dynamics
-- ISM: Structure}

\section{Introduction}

Hot, massive stars have
a major impact on the surrounding interstellar medium (ISM), not only at the
end of their life when they become supernovae but also throughout their
more stable phases of evolution via their strong stellar wind.
The winds of O-type stars can inject as much energy into the ISM
throughout their main sequence life as in their final explosion, and
therefore should have an equally great impact on the structures in the
ISM and its energization.

Typically, wind-blown shells have been found by looking at the
environments of stars known to have strong stellar winds, e.g.\ O stars
and Wolf-Rayet stars (e.g.\ Benaglia \& Cappa 1999, Marston 1997,
Miller \& Chu 1993, Dubner 
et al.\ 1990), but a neutral shell may continue 
to exist after the central star has evolved off the main sequence
and has lost the power to ionize the shell. 
Supernova remnants (SNRs) are thought to be visible in the radio
continuum for only a few to several tens of thousands of years (Braun
et al.\ 1989, Frail et al.\ 1994). As this is considerably
less than the duration of their life before they merge with the ambient
ISM ($> 1$ Myr), there should be many SNRs consisting of
neutral gas shells.
Large scale, low
resolution surveys have revealed the presence of \hi supershells
(e.g.\ Heiles 1984), most likely created by stellar clusters and
associations, but could not bring to light the smaller but likely numerous
neutral shells created by single stars which are an important part of
the Galaxy's zoology. 

The Canadian Galactic Plane Survey (CGPS; Higgs 1999 and Taylor 1999)
offers the first 
opportunity to study a large collection of \hi
shells, as opposed to supershells, to determine their dynamics and how
they relate to and impact on the surrounding ISM. 
A few such objects were serendipitously discovered in the pilot
project (Normandeau et al.\ 1997; hereafter NTD97).
One of these, \g132, will be presented in
detail here as an illustrative case study of this class of objects. It
is a striking feature within the \hi data cube, developing over
several spectral channels, at velocities generally associated with
interarm gas. 

The following section briefly outlines the observations and
processing of the data. Section 3 provides a description of the
structure at several wavelengths. In \S4, the possible location of
the shell is discussed. The next section considers the stars
present in this vicinity as possible energy sources for a wind-blown
bubble. In \S6, all these 
elements are brought 
together for analysis and conjecture. A summary and conclusions are
given in \S7.

\section{The data: observations and processing}

Radio continuum data at 408 MHz and 1420 MHz
as well as 21 cm spectral line data
were obtained at the Dominion Radio Astrophysical
Observatory (DRAO) as part the CGPS pilot project. The
pilot project covered an 8\arcdeg $\times$ 6\arcdeg\ area of the sky,
encompassing 
all of the W3/W4/W5/HB3 Galactic complex. Observations
were carried out in June, July, November and December of 1993.
Details of observations and data reduction are given in NTD97 except
for the 1420 MHz 
continuum polarisation data which are treated by Gray et al.\
(1999). Table \ref{tb:obs} 
summarizes the observational parameters for the 
DRAO data.

The CGPS also comprises other data sets which have been reprojected
and regridded to match the DRAO images. Among them is the FCRAO CO
Survey of the Outer Galaxy which is described by Heyer et al.\ (1998).

\section{Description}

\subsection{The \hi structure}

\label{sec:HI_descrip}

In the \hi images at velocities of approximately --25 \kms there is a
well-defined ring of enhanced emission, presumably a shell of atomic
hydrogen.  This shell is centred at $(l,b) = (132.62\arcdeg,
-0.72\arcdeg)$ and will henceforth be referred to as
G132.6--0.7--25.3.  Figure~\ref{fig:my_shell} presents a subsection of
the \hi mosaics for the relevant velocity interval.

For a complete expanding shell, the varying line-of-sight component of the
expansion velocity from projected centre to rim will result in the
constant velocity images in a data cube showing a progression from a
small filled ellipse (the receding cap), through annuli of
progressively larger radii, then decreasing radii back to a small
filled ellipse (the approaching cap).  \g132 appears to develop from a
cap at $v = -15.37 \kms$ to a complete ring at --25.27 \kms.  As
velocities become more negative it does not progress back to a cap.

At maximum extent \g132 is approximately elliptical.  The major axis,
which is perpendicular to the Galactic plane, measures 110.4 arcmin
and the minor axis is 95.2 arcmin; this is equivalent to 71 pc by 61
pc for a distance of 2.2 kpc (but see \S\ref{sec:distances} for a
discussion of possible distances). There is a hint that the structure
is slightly ovoid, being wider nearer to the plane.  For the half of
the shell from $-15.37 \kms$ to $-$25.27 \kms, the total flux above
the background level is $390\pm10\ {\rm Jy}$.

Within the shell at velocities where it is at full extent there is a 
\hi filament. At --25.27 \kms it diagonally traverses most of 
the shell. At more negative velocities, it persists along with a section
of the western edge of the shell, forming a U-shaped structure. It is
centred on ($132.3\arcdeg$, $-0.8\arcdeg$).

\subsection{Counterparts at other frequencies}

\subsubsection{Radio continuum}

Figure~\ref{fig:G132.6--0.7_cont} shows the
corresponding area from mosaics of the radio continuum emission at 408
MHz and 1420 MHz.
There is no corresponding ring structure in the radio continuum at
either frequency.

Figure~\ref{fig:pol} shows images of polarized radio emission from the
region of the shell. The emission is displayed in two equivalent forms:
Stokes Q and U images, and polarized intensity and polarization angle.
Highly structured 
emission is seen within \g132.  Outside of the ring, to the
northeast and northwest the polarised intensity vanishes.  

Gray et al. (1999) discuss the observations of the polarised 
emission from this region.  Polarized structures on arcminute
to degree scales are shown to arise from line-of-sight
variations in Faraday rotation of the diffuse Galactic
synchrotron radiation field.  The Faraday screen of varying
magentic field strength and ionized gas density is 
located primarily in the diffuse interstellar medium of
the Perseus arm.  The emergent radiation exhibits angular
structure in the polarisation angle of the polarised component.
The area to the north of \g132 is depolarized
due to the high electron density (and thus Rotation Measure)
in the ionized halo of the W3/W4 HII region complex (Gray et al. 1999).  
The appearance of polarised structures along lines of
sight within the ring suggests that the bubble is isolated
from the depolarising effects of W3/W4, perhaps because of the
surrounding protective shell of neutral gas at the rim.

It is noteworthy that the polarised emission shows structures
that are elongated in the northeast-southwest direction and
coincident with the HI filament that crosses the centre of the
bubble (see Figures 1 and 6).  This similarity suggests either
a diffuse electron component mixed in with the HI filament or
that magnetic fields play a role in the structure of this 
low-density environment.

\subsubsection{Infrared}

The IRAS infrared data have been searched for dust counterparts to the
atomic hydrogen structure. The 60 $\mu$m and 100 $\mu$m images of this
area are shown in
Figure~\ref{fig:G132.6--0.7_IRAS}.  Within the shell, in projection,
there is a clumpy 
plateau of infrared emission at 60 microns and 100 microns which falls
off rapidly at the HI boundary. 

\subsubsection{Molecular} 

Within the velocity interval and region of the shell there are no
extended molecular gas structures present in the FCRAO CO Survey of
the Outer Galaxy.  There are relatively compact sources projected onto
the eastern rim of the shell at ($133.19\arcdeg$, $-0.34\arcdeg$) and
($133.23\arcdeg$, $-0.62\arcdeg$) at velocities of approximately
--25.2 \kms. There is also a compact molecular cloud in the lower rim
of the shell at ($132.47\arcdeg$, $-1.37\arcdeg$, --28.5 \kms). The
velocities suggest that these are not merely along the line-of-sight
towards the rim but are in fact within it.

\subsection{Summary of morphology}

It can always be argued that any ``object'' seen in \hi images is but
a chance superposition of unrelated regions of emission however in
this case the accumulated evidence is reassuring.
The gradual progression from cap to full
extent in the \hi images,
the fall-off of infrared emission outside the shell (except to the
northeast), and the signature in the polarisation images, particularly
in the polarised intensity, of a clear difference between inside and
outside the shell all combine to show that \g132 is indeed a single, coherent
structure.

\section{Distance}

\label{sec:distances}

Assigning a distance or even a relative position along the
line-of-sight to \g132 is not an easy task. Different
possibilities emerge depending on the observational facts considered and
the assumptions made.

From the average velocity-longitude plot in Fig.~\ref{fig:vlplots}
(top panel) the shell's velocity would place it in the interarm region if
each of the main bands of emission is identified with an arm.
However the \hi distribution varies significantly over the latitude
range covered by the pilot project.  Judging from the
velocity-longitude plot for $b = -1.0\arcdeg$, the shell would be
at the outer edge of the Local \hi.
As for kinematic distances, the shell is completely developed at
$-25.27 \kms$.  Assuming a flat rotation curve with $A = 14 \kms {\rm
kpc}^{-1}$ and R$_0$ = 8.5 kpc, one finds d$_{\rm kin}$ = 1.7 kpc
(Burton 1988).  Using the best fit rotation curve from Fich et al.\
(1989), one finds 2.0 kpc. 

As was mentioned in the previous section, \g132 develops from a cap to full
extent over a range of velocities but does not progress back to a cap,
i.e.\ only half a shell is seen.
The missing second half implies that it must be on a density
gradient along the line-of-sight.
The location of \g132 on the edge of a spiral arm could account for
the absence of the second half: it would have expanded
more freely in this direction and would have fragmented and dispersed
into the less dense medium.  In this context, if the shell is now
static, the missing second half indicates that the shell is on the
outer edge of the Local arm.  However, if the shell is expanding, then
the less negative velocities correspond to material which is moving
away from us, and the suggestion is then that \g132 is on the near
side of the Perseus arm.  Alternately, the missing second half of the
shell could be indicative of the expansion in that direction having
been forestalled when it encountered a region of higher density. If
this is the case then the shell would either be near the edge of the
Local arm, the higher density of which would have halted the expansion
in that direction, or just past the high-density shocked region of the
Perseus arm. A cap would not be visible in this scenario because the
material in the second half of the shell would be indistinguishable
from the ``wall'' of higher density material which
prevented its expansion.

There is absorption associated with the
W3 \hii out to $-50 \kms$ (see e.g.\ Normandeau 1999), approximately
twice the velocity of the shell at 
full extent. If both the 
shell and the absorbing gas at --50 \kms are following the
rotation curve of the Galaxy then
W3 is further away and therefore
\g132 is at a distance of substantially less than 2.2 kpc, the
distance adopted here for W3. If, on the
other hand, the gas producing the absorption at --50 \kms is the shocked
gas prescribed by the Two-Armed Spiral Shock model (TASS; Roberts 1972) then
the shell is slightly further than W3, 
assuming that it is following the rotation of the Galaxy along
with the gas at the position of W3 at these velocities. In the TASS
model, the high density gas in the Perseus arm has been accelerated
from its standard rotation curve velocity of
approximately --20 \kms, and gas at --25 \kms or so would be
undisturbed gas that is located a little farther than the shocked gas;
the presence of emission rather than absorption at --25 \kms in the spectrum
towards W3 supports this idea (Normandeau 1999).

At the velocity where the shell is most clearly seen there is also 
interaction between the western edge of the W5 \hii region and the \hi, and 
there is \hi apparently associated with HB3 from --25.27 \kms to --28.00 \kms
and perhaps at --30.21 \kms (see NTD97).
If this apparently interacting \hi and 
the \hi forming the shell
is all at the distance of W5 and HB3 
then \g132 would be at $\sim$2.2 kpc. 

Table~\ref{tb:distances} summarizes this rather confusing state of
affairs. 
In what follows, all quantities shall be given with their
dependance on distance expressely stated and with the value for a
distance of 2.2 kpc in brackets. 
This value is preferred for a combination of reasons.
Kinematic distances have shown themselves to be unreliable towards the
Perseus arm (eliminating entries {\small I} and {\small II} of the
table), tied to the fact that the standard rotation curve does not
apply because of observed streaming motions (eliminating entry {\small
IV}, as well as {\small VI} and 
{\small IX} both of which implicitely assume that all the \hi is
following the rotation curve, that decreasing velocity corresponds to
increasing distance). The TASS model is a more promising description
of the behaviour of gas towards these longitudes, favoring entry
{\small V} (slightly more than 2.2 kpc) which places \g132 slightly
past the main ridge of the Perseus arm (entry {\small XI}). This is
also in accord with the inference that it is at the same distance as
W5 and HB3 (entry {\small III}).

\section{Stars}

\label{sec:stars}

There are no visible, catalogued, energetic main-sequence stars within
the shell at present (according to the Simbad data base).  
This is not surprising; if there were energetic stars present, there
should be an inner shell of ionized gas visible in the Stokes I images.
Figure \ref{fig:G132.6--0.7_stars} shows the positions of the 74
catalogued O and B stars in the vicinity (in projection) of \g132 with
reference to the shell at full extent. The area searched using the Simbad
data base was the one displayed in the figure. The concentration of
stars in the 
lower left-hand corner of the plot is the open cluster Stock 2 which
is at 303 pc (Mermillod 1999). A cautionary note should be sounded: if
\g132 is behind the main ridge of the Perseus arm as 
argued above, then some stars may have been lost to
obscuration. However, at these longitudes the plane is at higher
latitudes, near $b = 1\arcdeg$, and most of the observed dust emission seems to
be associated with the shell rather than being in the foreground.  

The most promising candidate for energy source of the shell if it is
wind-blown is BD+60 447.
This B1 Ia star is almost exactly at the centre of the shell in
projection, and 
according to Humphreys (1970) it is at a distance of 1.55 kpc,
determined spectro-photometrically from previously published data,
which distance is not inconsistent with the various 
estimates for \g132.
No uncertainty was quoted by Humphreys and there was no radial velocity 
listed. While on the main sequence BD+60 447 was most probably a late O star,
which means that it would have had sufficiently strong stellar winds
to blow a bubble even though, in its present state, it is no longer
capable of maintaining the growth of the shell or its ionization. 

Other stars within the shell in projection include main sequence B9
and B7 stars, and four unclassified B stars. While these do not
provide stellar winds, if they are inside the shell they
may be contributing to the expansion of \g132 through radiation
pressure as discussed by Elmegreen \& Chiang (1982). These authors
contend that once a shell has grown sufficiently that it includes many
field stars, their radiation pressure will cause the shell's expansion
to accelerate.

Based on the evolutionary tracks by Maeder (1990) and using the
luminosity and effective temperature given by Lang (1991) for a B1 I star,
it would appear that BD+60 447 had a mass between 20 \Msun\ and 25 \Msun\ while
on the main sequence. According to Table 3 of Howarth \& Prinja
(1989; hereafter HP89)
this implies that it was an O9.5 V or an O9.0 V star. 
As a lower limit for
the energy that could have been input by stellar winds during its hydrogen
burning phase, consider an O9.5 V star. 
Based on the empirical relations derived by HP89,
such a star would have a mass-loss rate of $10^{-7.36}$ \Msun
yr$^{-1}$ and a terminal wind velocity  
of 2000 \kms, giving a stellar wind luminosity of $6 \times
10^{34}$ erg s$^{-1}$. From Stothers (1972),  
the main sequence lifetime of such a star would be 11.2 Myr and therefore,
the total 
kinetic energy output would be $\sim 2 \times 10^{49}$ erg. 

\section{Analysis and conjecture}

It will be assumed that the shell is expanding. This is the most likely 
scenario in view of the varying morphology seen in the \hi images. It is
unlikely that there exists, in the ISM, 
a long enough, stationary cylinder or funnel 
to account for the aspect of the \hi in the different channels.

\subsection{Kinetic energy of \g132}

From its integrated flux, from which a twisted plane background was
subtracted, the average column density for
the well-defined first half of the shell is
$2.7 \times 10^{20}\ {\rm cm}^{-2}$.
This implies an \hi mass of $1.8\, d_{\rm kpc}^2 \times 10^3\ \Msun$
[$9 \times 10^3\ \Msun$].
This is in agreement
with Heiles (1984)'s statistical
observation that the mass swept up by a shell is 
very approximately $8.5 R_{\rm sh}^2\ \Msun$, where $R_{sh}$ is in
parsecs, which in this case would predict  
${\rm M} \sim1.9\, d_{\rm kpc}^2 \times 10^3\ \Msun$ [$\sim 9 \times
10^3\ \Msun$].

Assuming that the cap of the shell is seen at a velocity of $-15.37 \kms$ and
that it has reached full extent at $-25.27 \kms$, an expansion
velocity of approximately 9.9 \kms is found. This is barely larger than the
turbulent velocity standardly assumed for the ISM, implying that the shell
should soon begin to dissipate into the ambient gas, though the low density of
the latter, as evidenced in the image of the shell at full extent,
will cause the process to be slower than in denser surroundings.
For a complete shell --- one with twice the mass of the fore
half --- to expand with this velocity would require the injection
of $\sim 2\, d_{\rm kpc}^2 \times 10^{48}$ erg [$\sim10^{49}$ erg].
Note that this is the same order of magnitude as the stellar wind kinetic
energy of BD+60 447 during its main sequence life. Also, it is
low though not unreasonable for a supernova remnant, especially
considering that it would have lost energy by now; $5\, d_{\rm kpc}^2
\times 10^{49}$ erg [$2 \times 10^{50}$ erg] may have been lost to
recombination if all the \hi currently associated with \g132 was
previously part of an ionized shell.

\subsection{Age and expansion velocity}

If the winds from BD+60 447 created and sustained the shell, then \g132
would be approximately the same age as the star. Given that fusion
stages past hydrogen burning are estimated to last 0.1 times as long
as the main sequence (Meynet et al.\ 1994), the age of the bubble
cannot be much greater than the main sequence lifetime of BD+60 447.
This was earlier estimated to be 11.2 Myr based on the assumption
that it was of type O9.5 when on the main sequence. It should be noted
that the shell's low expansion velocity and
the current evolutionary phase of the central star are in accord with
the fact that, in the standard model for wind-blown bubbles (Weaver et al.\
1977), the time to dissipation into the ISM 
is approximately equal to the main sequence lifetime of the
source of the wind.

The shell should at present be in the momentum driven bubble phase,
but the transition to this phase would only have been a recent event
and therefore it would be best to consider the previous phase, a
bubble with a radiative outer shock.
According to the standard model, for a bubble
with a radiative outer shock the radius varies as 
\begin{equation}
R_2(t) = 28 \left(\frac{L_{36}}{n_0}\right)^{1/5} t_{6}^{3/5}\ {\rm pc},
\label{eq:R_shell_2}
\end{equation}
where $L_{36}$ is the wind luminosity in units of $10^{36}$ erg
s$^{-1}$ and the  
time is given in units of $10^6\ {\rm yr}$. 

By taking the derivative of the above equation and using the radius and age
estimates, one can calculate an expansion velocity for the
shell. 
A velocity of $0.7\, d_{\rm kpc}$ \kms
[1.6 \kms] is predicted and the shell would only have slowed down
further as it continued into the momentum driven bubble phase, barring
other energy inputs. 
Not only is this significantly less than the
observed value but it is also less than the 
turbulent velocity of the ISM and so the shell should have dissipated.
This shell has too high an expansion velocity for its radius and
assumed age if it is wind-blown. 

The kinematic age of the bubble ($R_{\rm sh} / v_{\rm exp}$), 
for a constant expansion velocity of 9.9 \kms, is $1.4\, d_{\rm kpc}$ Myr
[3.1 Myr]. This should be an upper limit to the age of the bubble,
unrelated to wind-blown models, as long as there has been no acceleration.
The age estimated from main sequence lifetime of the assumed stellar
wind source is 
much greater than the kinematic age stated above. BD+60 447 may have
been of a somewhat earlier type when on the main sequence, perhaps as
early as an O8, but this does not solve the problem because the age
estimate would still be too high, slightly greater than 7.1 Myr
(Stothers et al.\ 1972) which would require an uncertainty of
over 100\% on the kinematic age in order for there to be
agreement. This is unlikely considering that there is little
uncertainty in the radius, and as for the expansion velocity, the
smooth variation of the morphology from channel to channel argues
against the estimate of 9.9 \kms being significantly off.

This age disagreement indicates that the stellar winds from this star
on its own cannot be responsible for the 
present state of \g132. 
If \g132 has been mainly formed by BD+60 447's wind,
some other factor must have caused it to accelerate.
Expansion into a density gradient
and consequent acceleration (see next section), may explain the 
discrepancy between the observed velocity and that predicted for expansion
into a uniform medium. Another possible contributing factor is the
radiation pressure from ordinary stars now within the shell, as
mentioned in \S\ref{sec:stars}.

For the SNR hypothesis, there is no candidate energy source and
therefore no presumed age for the \g132. Thus
the kinematic age and the expansion velocity do not pose a problem if
the shell is a SNR rather than a 
stellar wind bubble. 

\subsection{Shape of the shell and scale height}

\label{sec:scale}

As stated in \S\ref{sec:HI_descrip}, \g132 is elongated in
the direction perpendicular to the Galactic plane, slightly wider at
the base. This type of shape is expected for a bubble evolving in a
density gradient. The minor axis is 95.2 arcmin (61 pc for 2.2 kpc). 
Since any model of bubble evolution in a stratified atmosphere
(e.g.\ Kompaneets 1960, Tomisaka \& Ikeuchi 1986, Mac Low \& McCray 1988)
predicts near spherical evolution at early times, and significant
elongation only when the radius exceeds the scale height (enabling
the bubble to sense the ambient stratification), the observed elongation
of this shell implies $H < b/2 = 13.8\, d_{\rm kpc}$ pc [30.5 pc].

This extremely small value for the scale height is reminiscent of the
low value of H (25 pc) found by Basu et al.\ (1999) for the nearby (in
projection at least) W4 superbubble, and of the scale height (22 pc) used by
Shelton et al.\ (1999) when modelling W44. 
A more precise value for the scale height can be found by fitting the 
analytic Kompaneets (1960) solution to the shell. The Kompaneets solution 
consists of an analytic expression for the bubble shape at various stages
of evolution in an exponential atmosphere. The observed ratio of major
to minor axis can be matched to a Kompaneets model at particular
stage of evolution, yielding the ratio of the current radius to the
ambient scale height. Details of the Kompaneets solution and this technique
for determining the scale height can be found in Basu et al.\ (1999).
For \g132, we find that the best fit Kompaneets
model has a semi-minor axis $1.76H$. 
This yields $H=7.9\, d_{\rm kpc}$ pc [17.3 pc]. The
general point is that the scale height must be 30 pc or less in this
environment, if the elongation is due to a density gradient.

This value is, of course, valid for the ISM local to \g132, just as
the values relating to studies of W44 and the W4 superbubble were
applicable to the environment of these objects. They do not invalidate
the much greater values ($>$ 100 pc) found for the global, Galactic
scale height, though this is perhaps an indication that the global
scale height is determined by very different processes than those
that govern the equilibrium of gas on smaller scales.

\subsection{Within the shell}

\label{sec:U}

As was noted in the description of the \hi emission, there is an \hi ``U''
within the shell. A partial shell within a shell as it were.
In the crook of this ``U'' there is a compact radio continuum source which is
positionally coincident with an
IRAS compact source having the colours of an \hii region
(see Hughes \& MacLoed 1989 for colour selection criteria), namely
IRAS-02044+6031. The infrared colours, the radio continuum spectral index of
+0.55 and the morphological indication of \hi surrounding the compact source
combine to suggest that this an \hii region within a layer of 
dissociated gas, located on the periphery of the shell.

Though originally thought to be a planetary nebula (Acker et al.\
1983), this identification of IRAS-02044+6031 has since been found to
be in error (Sabbadin 1986, Acker et al.\ 1986, Zijlstra et al
1990). No maser emission has been detected despite several searches
(6.7 GHz methanol by MacLoed et al.\ 1998, 5 cm OH lines by Baudry et
al.\ 1997, H$_2$O maser lines
by Codella et al.\ 1996 and by Brand et al.\ 1994), indicating that
either the geometry is simply inappropriate for maser detection or that the
region has evolved sufficiently that there is no longer maser
activity. Considering the possibility that the \hi $U$ is related
dissociated gas, the latter explanation is not unreasonable. However, it should
be noted that the 
velocity interval sampled by the maser searches did not always cover
velocities as high as 
for the \hi discussed here.

There is no indication of a bright star which could account for the ionized
gas. In fact on the POSS images, coincident with the IRAS source,
there is a compact region of
increased extinction. This suggests that this is a young \hii region. Perhaps
its formation was triggered by the expansion of the \g132 shell. It
should be noted however that Wouterloot \& Brand (1989) associate this
IRAS source with CO emission at --55.7 \kms which would be unlikely to
be related to the \hi seen at --25.27 \kms, and the geometry of the
region of CO emission at --55 \kms is also not suggestive of an
association with the \hi.

To summarize,
in the context of the larger shell described in this paper, this small
U-shaped structure is proposed to be \hi formed by the stars within a
compact \hii 
region through dissociation. The \hii region itself was perhaps
formed when the expansion of \g132 compressed gas in its
periphery sufficiently to induce star formation. The U would then be
second generation \hi gas related to \g132. 

\section{Summary and conclusions}

An \hi shell has been found near the very active W3/4 \hii
region complex. The lack of a radio continuum counterpart has been
interpreted as indicative of the advanced age of the shell,
be it a wind-blown shell or a SNR. 
If it is a wind-blown shell then the most likely powering source is 
the B1 supergiant BD+60 447. This is based on the position of the star (at
the centre in projection and at a reasonable distance) and its spectral
type (strong enough stellar winds while on the main sequence to blow a
bubble, no longer capable of maintaining the ionization of the shell
which is in accord with the lack of a continuum counterpart). The age
of the star and the kinematic age of the shell are however discrepant,
the former being greater than the latter. This could point to the
shell being a member of the observed class of ``high velocity''
shells (Oey 1996), which have somehow been reaccelerated, 
but it could also be taken to indicate that the shell was not 
created by BD+60 447 (and there are no other catalogued stars present
which could have blown the shell through stellar winds) but is in fact
a SNR. Based on the available data, it is not 
possible to distinguish between the two possibilities.

Regardless of the origin of the \g132, if the elongation of the shell
is due to the density gradient of its surroundings, the Kompaneets
model can be 
used to determine the scale height of the ambient ISM. In this case,
the aspect ratio of the shell indicates that the local scale height is 17.3
pc for a distance of 2.2 kpc. While surprisingly low, analysis of
other regions have also pointed to small scale heights (W4 by Basu et
al.\ 1999; W44 by Shelton et al.\ 1999).

At the edge of the shell, there is a smaller U-shaped \hi structure
curving around a compact radio continuum and infrared source. The thermal
spectral index of the compact source and its infrared colours imply
that it is an \hii region and the \hi then corresponds to an
encircling photodissociation region. However, the FCRAO outer Galaxy
Survey shows no indication of a coincident molecular cloud at similar
velocities. It has been hypothesized that this \hii region could be
the result of star formation triggered by the expansion of
\g132. Compact $^{12}$CO clouds at other locations along the shell's
perimeter could also be triggered or enhanced condensations. 

The data from the CGPS are likely to be rife with such structures as
their identification requires arcminute resolution coupled with
coverage of wide angular scales. With an analysis of this sort carried
out for each shell, our picture of the star-ISM feedback mechanisms
will be more complete. 

\acknowledgements

M.N.\ thanks Brad Wallace for useful input, as well as James Graham
and Carl Heiles for comments on drafts of this paper. 
The Dominion Radio Astrophysical
Observatory's synthesis telescope is operated by the National Research
Council of Canada as a national facility. 
The Canadian Galactic
Plane Survey is a Canadian project with international partners, and is
supported by a grant from the Natural Sciences and Engineering Research
Council of Canada.  
This research made extensive use of the Simbad database, operated at CDS, 
Strasbourg, France, and of NASA's Astrophysics Data System Astrophysics
Science Information and Abstract Service.

\clearpage

\begin{deluxetable}{l l l}
\tablewidth{0pt}
\tablecaption{Observational parameters for the synthesis telescope data
\label{tb:obs}}
\tablehead{
\colhead{Parameter} & \colhead{Frequency} & \colhead{Value} 
}
\startdata
Bandwidth & 408 MHz & 4 MHz \nl
 & 1420 MHz continuum & 30 MHz \nl
 & 1420 MHz spectral line & 1 MHz \nl
Polarisation & 408 MHz & RR \nl
 & 1420 MHz continuum & RR, LL, RL, LR \nl
 & 1420 MHz spectral line & RR \nl
Spatial resolution & 408 MHz & $3.5' \times 4.0'$ (EW $\times$ NS) \nl
 & 1420 MHz & $1.00' \times 1.14'$ (EW $\times$ NS) \nl
Central velocity (LSR) & 1420 MHz spectral line & --50.0 \kms \nl
Velocity coverage & 1420 MHz spectral line & 211 \kms \nl
Channel width & 1420 MHz spectral line & 2.64 \kms \nl
Channel separation & 1420 MHz spectral line & 1.649 \kms \nl
Sensitivity & 408 MHz & 1.9 mJy/beam --- 16.5 mJy/beam \nl
 (theoretical) & 1420 MHz continuum & 0.23 mJy/beam --- 1.15 mJy/beam \nl
 & 1420 MHz spectral line & 3.0 K --- 15.0 K \nl
\enddata
\end{deluxetable}

\clearpage

\begin{deluxetable}{l l l}
\tablecaption{Possible distances and relative positions
\label{tb:distances}}
\tablehead{
\colhead{} & \colhead{Distance or position} & \colhead{Basis$^*$}
}
\startdata
{\small I} & 1.7 kpc & Flat rotation curve \nl
{\small II} & 2.0 kpc & Fich et al.\ (1989) \nl
{\small III} & $\sim$2.2 kpc & \hi interacting with W5 and with HB3 \nl
{\small IV} & $\ll 2.2$ kpc & W3 absorption and standard rotation curve \nl
{\small V} & 2.2+ kpc & W3 absorption and TASS model \nl
{\small VI} & edge of Local arm & $v-l$ plot for $b = -1.0\arcdeg$ \nl
{\small VII} & edge of Local arm & static shell, missing second half dispersed \nl
{\small VIII} & edge of Local arm & expanding shell, missing second half stalled \nl
{\small IX} & interarm & $v-l$ plot for all $b$ \nl
{\small X} & near side Perseus arm & expanding shell, missing second half dispersed \nl
{\small XI} & behind main ridge of Perseus & expanding shell, missing second half stalled \nl
\enddata
\tablenotetext{*}{See text (\S\ref{sec:distances}) for explanations}
\end{deluxetable}

\clearpage

\figcaption[figures/shell_HI.ps]{The \hi shell \g132.
	Greyscales vary linearly between 15 to 80 K. Velocities with
	respect to the Local Standard of Rest are indicated at the top
	left of each panel.
	\label{fig:my_shell}}

\figcaption[figures/G132607_cont.ps]{The radio continuum in the \g132 region.
        The left panel shows the 408 MHz continuum emission and the
        1420 MHz image is on the right. Greyscales vary linearly from 
        10 K to 150 K for the 408 MHz image and from 0 K to 5 K for the
        1420 MHz image.
        A $110.4 \times 95.2\ {\rm arcmin}^2$ ellipse corresponding roughly 
        to the shape of the shell at full extent is overlaid.
	\label{fig:G132.6--0.7_cont}}

\figcaption[figure/G132607_pol.ps]{1420 MHz polarisation images. Top
	left: Stokes Q, greyscale is from --0.25 to +0.25 K. Top
	right: Stokes U, greyscale is from --0.25 to +0.25 K. Bottom
	left: polarized intensity, greyscale is from 0.0 to 0.35
	K.. Bottom right: polarisation angle, greyscale is from
	--90\arcdeg to +90\arcdeg. A $110.4 \times 95.2\ {\rm
	arcmin}^2$ ellipse corresponding roughly  
        to the shape of the shell at full extent is overlaid. 
	\label{fig:pol}}

\figcaption[figures/G132607_IRAS.ps]{The 
	 infrared emission in the \g132 region. 
	 The 100 $\mu$m image is on the left and the 
	 60 $\mu$m on the right. Greyscales vary linearly from 45 MJy/sr
	 to 120 MJy/sr for the 100 $\mu$m image and from 6 MJy/sr to 36 MJy/sr 
	 for the 60 $\mu$m image.
	 A $110.4 \times 95.2\ {\rm arcmin}^2$ ellipse corresponding roughly 
	 to the shape of the shell at full extent is overlaid. The
	 bright compact source at $l = 132.16$ and $b = -0.72$, surrounded by
	 obvious processing artifacts, is IRAS-02044+6031 which is
	 discussed in \S\ref{sec:U}.
	 \label{fig:G132.6--0.7_IRAS}}

\figcaption[figures/v-l_plots.ps]{Longitude-velocity plots for the
	CGPS pilot region. The greyscale varies from white at 0 K to
	black at 80 K. The top panel is an
	average over all latitudes and the second panel is for $b =
	-1.0\arcdeg$. 
	\label{fig:vlplots}}

\figcaption[figures/ch50_OBstars.ps]{O and B stars in the
        projected vicinity of the \g132.
        Open star symbols are for B type stars and filled star symbols
        are for O type stars. The greyscale 
        showing the \hi at --25.27 \kms varies linearly between 15 and 80 K.
	\label{fig:G132.6--0.7_stars}}

\clearpage

\begin{figure}
\vspace{20cm}
\includegraphics{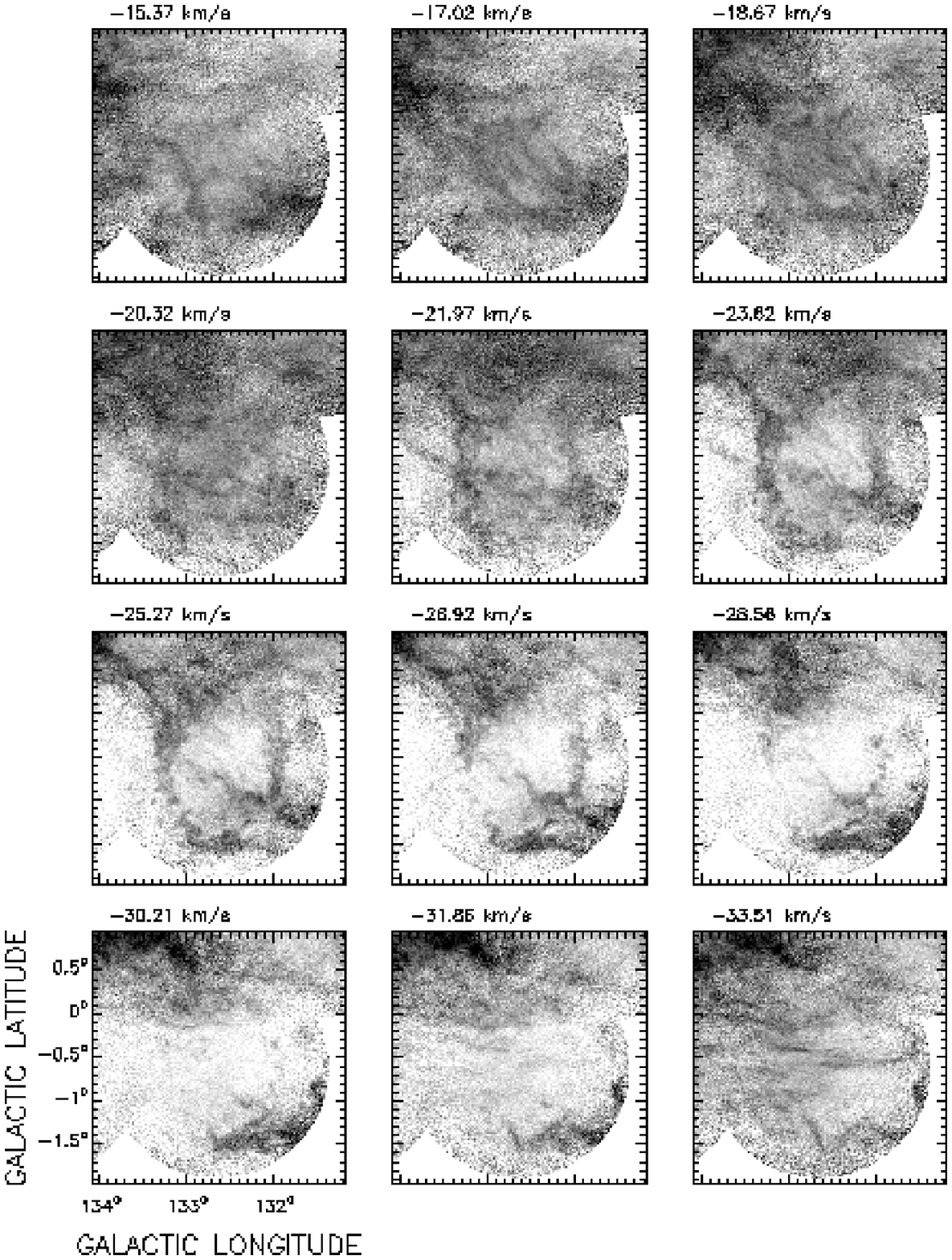}
\end{figure}

\clearpage

\begin{figure}[t]
\vspace{8cm}
\includegraphics{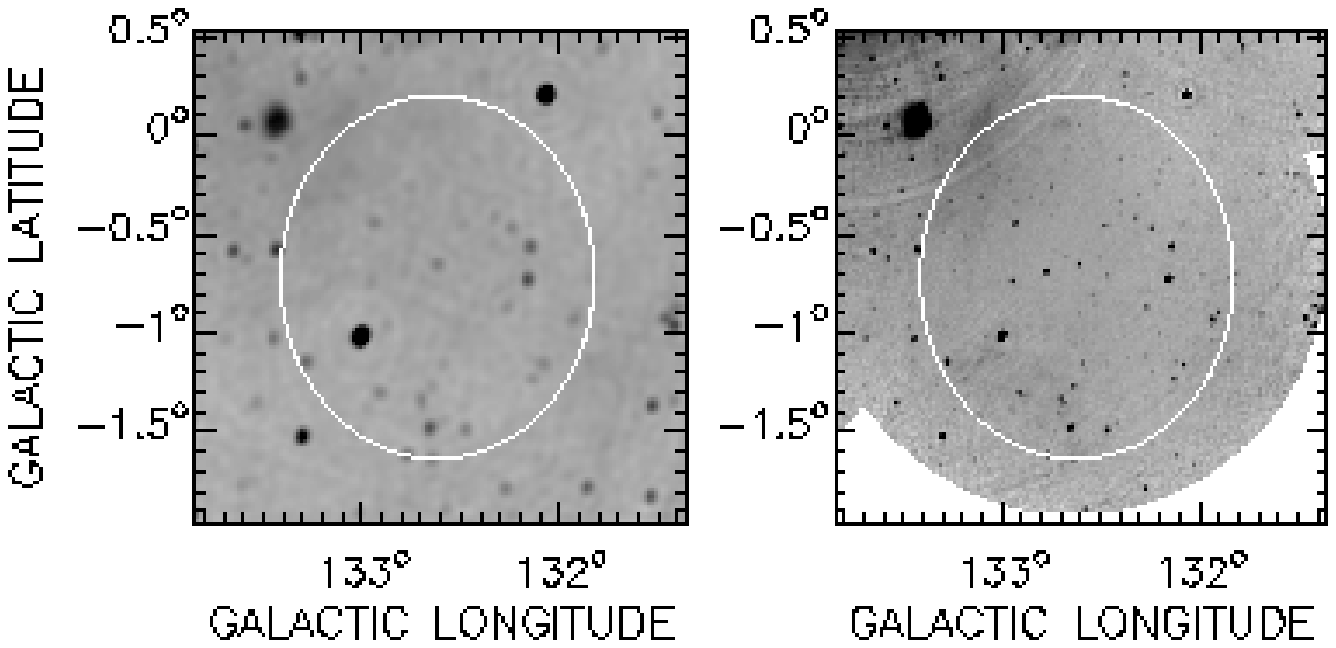}
\end{figure}

\clearpage

\begin{figure}[t]
\vspace{15cm}
\includegraphics{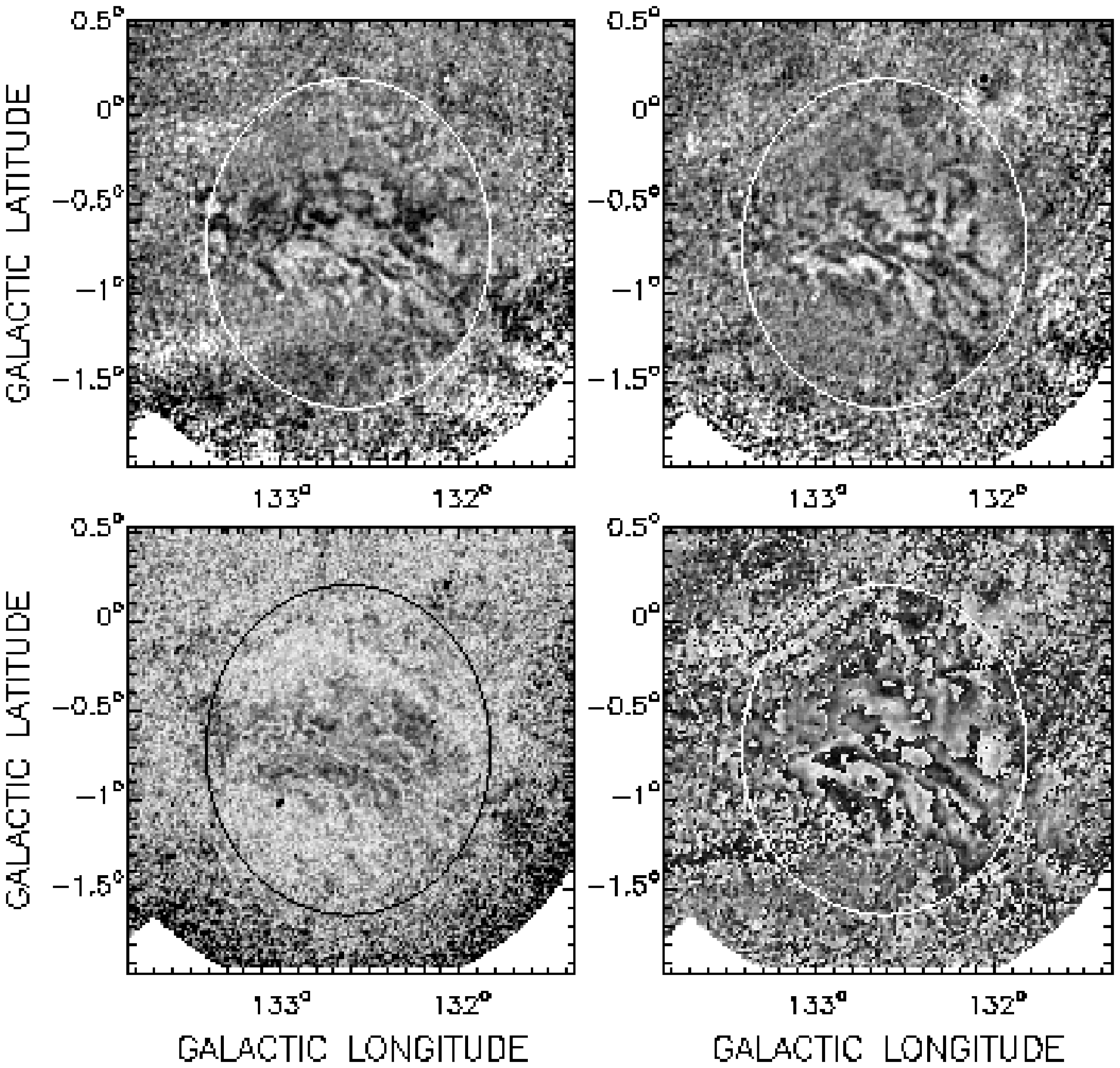}
\end{figure}

\clearpage

\begin{figure}[t]
\vspace{8cm}
\includegraphics{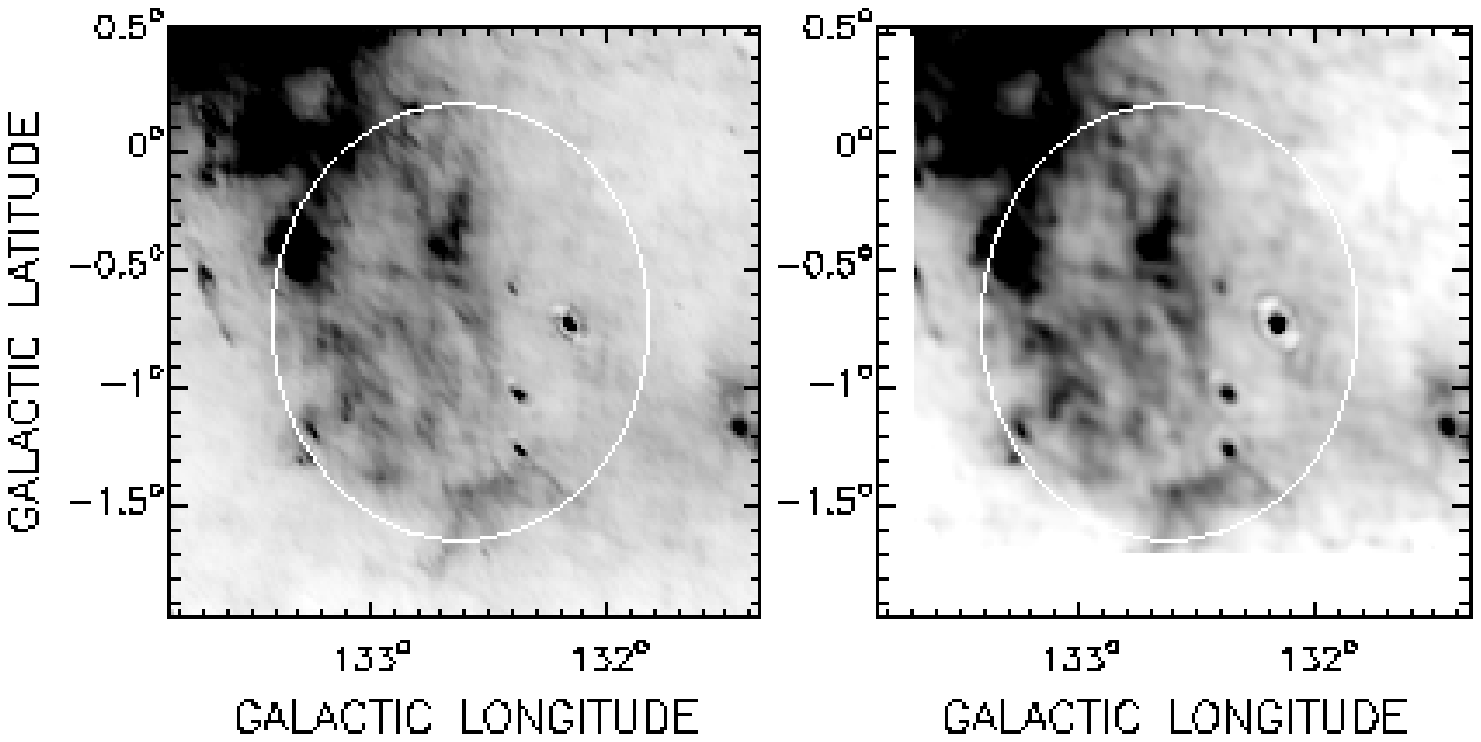}
\end{figure}

\clearpage

\begin{figure}[t]
\vspace{10cm}
\includegraphics{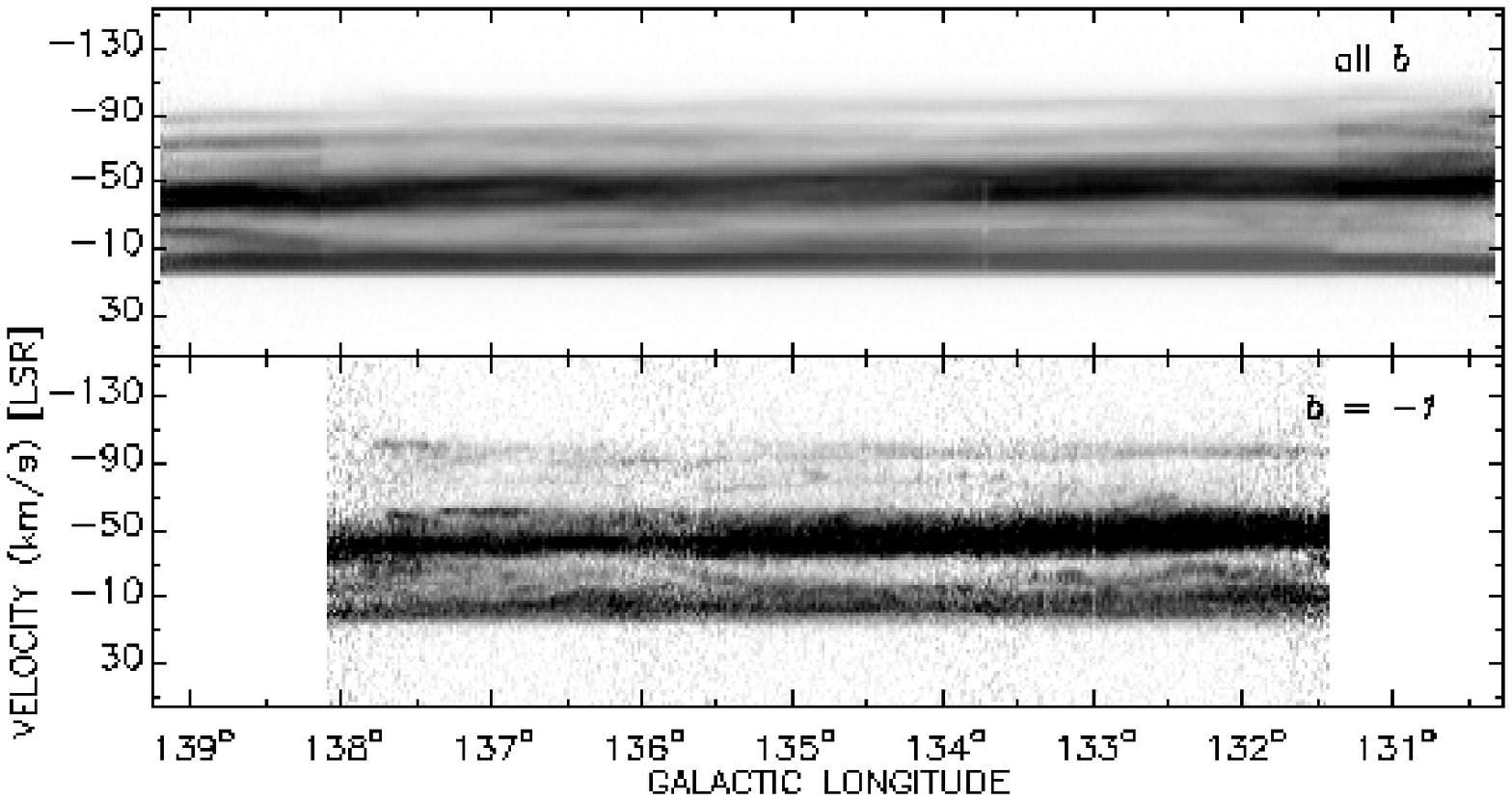}
\end{figure}

\clearpage

\begin{figure}[t]
\vspace{8cm}
\includegraphics{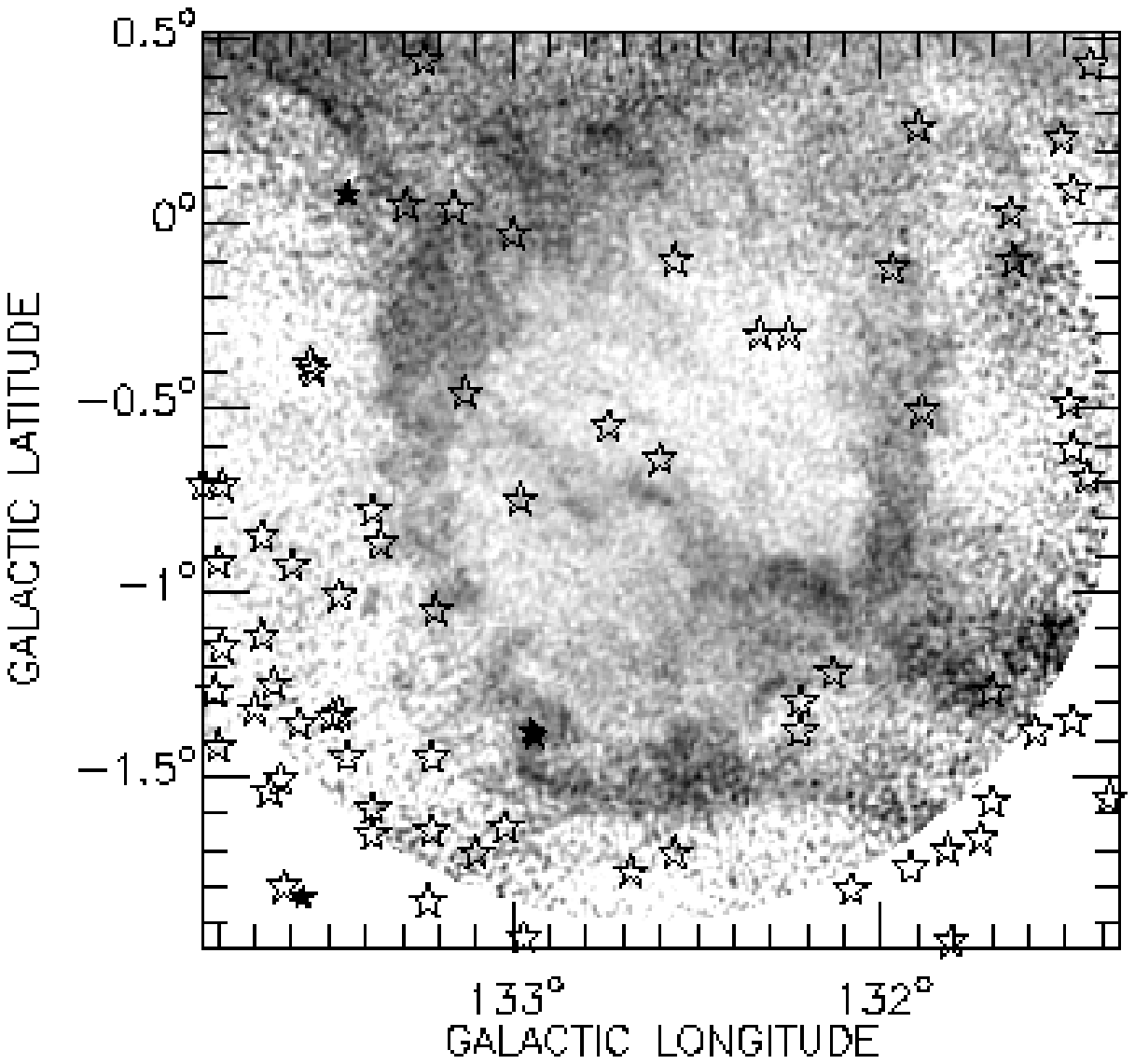}
\end{figure}


\end{document}